\documentclass[preprint,3p]{elsarticle}
\usepackage[scaled=0.99]{helvet}
\usepackage[T1]{fontenc}
\usepackage[latin9]{inputenc}
\usepackage{mathtools}
\usepackage{amsmath}
\usepackage{amssymb}
\usepackage{graphicx}

\makeatletter

\usepackage{epstopdf}

\usepackage{url}
\usepackage{natbib}
\biboptions{sort&compress}

\usepackage{amsfonts}\usepackage{latexsym}
\usepackage{amsthm}\usepackage{amsfonts}\usepackage{mathrsfs}\usepackage{bbm}\usepackage{enumerate}
\usepackage{color}
\usepackage{pst-func}
\usepackage{dcolumn}
\usepackage{bm}

\renewcommand{\epsilon}{\varepsilon}

\makeatother

\begin{document}

\title{ Breakdown of the Fermi polaron description near Fermi degeneracy at unitarity }

\author[swin]{Brendan C. Mulkerin}

\author[swin]{Xia-Ji Liu}

\author[swin]{Hui Hu}

\address[swin]{Centre for Quantum and Optical Science, Swinburne University of Technology,
Melbourne 3122, Australia.}


\date{\today}
\begin{abstract}
We theoretically investigate attractive and repulsive Fermi polarons
in three dimensions at finite temperature and impurity concentration
through the many-body $T$-matrix theory and high-temperature virial
expansion. By using the analytically continued impurity Green's function,
we calculate the direct rf spectroscopy of attractive polarons in
the unitary regime. Taking the peak value of the rf spectroscopy as
the polaron energy and the full width half maximum as the polaron
lifetime, we determine the temperature range of validity for the quasi-particle
description of Fermi polarons in the unitary limit.  
\end{abstract}

\maketitle

\section{Introduction}

Understanding and exploring the properties of a moving impurity immersed
in quantum many-body systems - the so-called polaron - is a fundamental
problem in condensed matter physics and ultracold physics \cite{Massignan2014}.
In particular, with the advent of highly controllable ultracold systems
\cite{chin2010}, the polaron problem has become a recent topic of
importance in both Fermi \cite{Schiro2009,nasc2009,Navon2010,Scazza2017}
and Bose gases \cite{jin2016,nils2016}. The use of magnetic Feshbach
resonances allows for the control of the interactions between the
impurity and the background system, and with the advancement of experimental
apparatus the dimension of the system can be readily changed from
three dimensions to two dimensions to explore the role of dimensionality
\cite{kohl2012,kohstall2012,Levinsen2015}.

The Fermi polaron constitutes the extreme case of a highly spin-imbalanced
Fermi gas and is the conceptually simplest strongly correlated many-body
system. It is thought to be a key to better understand imbalanced
strongly interacting Fermi mixtures \cite{nasc2009,Chevy2010,Schmidt2018}
at the crossover from a Bose-Einstein condensate (BEC) to a Bardeen-Cooper-Schrieffer
(BCS) superfluid \cite{Bloch2008,Chevy2010}. Initial experimental
work of Fermi polarons focused on the attractive branch \cite{Schiro2009},
however there exists a higher metastable state when there is a weak
two-body molecular state, known as the repulsive polaron \cite{Massignan2008PRA,cui2010}.
The repulsive polaron in a three-dimensional two-component Fermi gas
has been recently explored in detail at the European Laboratory for
Non-Linear Spectroscopy (LENS), Florence \cite{Scazza2017}. Probing
the repulsive Fermi polaron may provide insight to understand repulsively
interacting many-body systems such as itinerant ferromagnetism \cite{Jo2009,Sanner2012,He2016PRA,Valtolina2017}.

Experimentally, both attractive and repulsive polarons have been probed
through the use of radio-frequency (rf) spectroscopy \cite{Chin2004,Shin:2008,Stewart2008},
which allows the measurement of the single-particle spectral function
of the polaron. The system is initially prepared with three spin states,
a majority $\left|\uparrow\right\rangle $ state, and two minority
states, $\left|i\right\rangle $ and $\left|f\right\rangle $, where
the interaction between the majority and each minority state can be
tuned. The rf spectroscopy works by applying a short rf pulse to flip
the impurities into a third state, and by varying the detuning of
the rf pulse, information about the single-particle excitations can
be measured. For \emph{direct} rf spectroscopy the third state has
been tuned to be non-interacting with the $\left|\uparrow\right\rangle $
state, the strongly interacting impurities are flipped into this non-interacting
state. \emph{Reverse} rf spectroscopy is the opposite mechanism, the
impurity-majority interaction is initially weak and the impurities
are flipped into a strongly interacting third state, which allows
for the full excitation spectrum to be more easily probed.

Theoretically, attractive and repulsive polarons have been extensively
studied at \emph{zero} temperature within a variational framework
for a wide range of Fermi gases \cite{chevy2006,lobo2006,combescot2007,Punk2009,mathy2011,levinsen2015pol},
Bose gases \cite{Sashi2014,li2014}, Fermi superfluids \cite{Nishida2015,Yi2015},
and long-range interacting systems \cite{Kain2014,Camargo2018}. Sophisticated
diagrammatic quantum Monte Carlo (QMC) schemes have been developed
to tackle the polaron problem \cite{prokefev2008,Vlietinck2014,kroiss2015,Goulko2016},
with excellent agreement between the QMC simulations and many-body
results. At finite but low temperature there have been few calculations
for either the Fermi or Bose polaron \cite{Massignan2008PRA,hui2017,Liu2018,Tajima2018,Guenther2018},
this may be in part due to the success of the variational ansatz in
describing the polaron systems.

In this work, we explore the properties of the Fermi polaron in three
dimensions at \emph{high} temperature close to Fermi degeneracy and
address the validity of the Fermi polaron description, as motivated
by the on-going measurement at Massachusetts Institute of Technology
(MIT) \cite{Struck2017}. Our investigation uses a many-body $T$-matrix
theory and a non-perturbative virial expansion theory \cite{Liu2009,Kaplan2011,Leyronas2011,Liu2013}.
The former is a diagrammatic approach that includes pair fluctuations
in the normal state and is known to well represent strongly interacting
Fermi gases in the limits of spin-balance \cite{Haussmann2007,Ohashi2009,Mulkerin2016}
and imbalance \cite{Veillette2008,massignan2008,Massignan2008PRA,Klimin2012,doggen2013}.
At zero temperature and for a single impurity, the many-body $T$-matrix
and the variational ansatz are known to be equivalent in three dimensions
\cite{combescot2008,bruun2010,Massignan2011,schmidt2011}. The latter
virial expansion theory works very well at high temperature above
the Fermi degenerate temperature.

It is expected that the quasi-particle and Fermi-liquid description
of the polaron will break down as the temperature of the system increases
and the Fermi surface broadens significantly due to thermal fluctuations
\cite{Sagi2014,Reichl2015,Struck2017}. Here, we probe the temperature
breakdown of the quasi-particle description in the unitary limit,
i.e. where the $s$-wave scattering length is infinite, by finding
the temperature dependence of the full width half maximum (FWHM) of
the rf spectra, which corresponds to the lifetime of the polaron.
For temperatures above the Fermi degenerate temperature we find through
the $T$-matrix and virial expansion theory a very broad peak in the
rf spectra and hence the breakdown of the quasi-particle description.
As the temperature lowers below the Fermi temperature the peak in
the spectra becomes narrow and the FWHM becomes smaller than the polaron
energy (i.e., peak position) for temperatures below $T\simeq0.8T_{{\rm F}}$,
where we will then have a defined quasi-particle.



The paper is set out as follows: in Sect.~\ref{sec:methods} we outline
the many-body $T$-matrix theory and virial expansion theory for the
attractive and repulsive Fermi polarons. We show how to determine
the quasi-particle properties of polarons, connecting the $T$-matrix method to experimental and previous theoretical results, and briefly explain the
calculation of rf spectroscopy. In Sect.~\ref{sec:hightemp} we explore
the high temperature rf spectroscopy of the attractive polaron in
the unitary limit and investigate the breakdown of the quasi-particle
description. Finally, in Sect.~\ref{sec:conc} we conclude with a discussion
of our results. \ref{app:tmatrix} shows the calculation of the 
many-body $T$-matrix formalism and \ref{app:thermo} and 
\ref{app:Green} are devoted to the details of the virial expansion theory.

\section{Methods}

\label{sec:methods}

We consider a highly spin-imbalanced two-component Fermi gas (i.e.
$n_{\uparrow}=n\gg n_{\downarrow}$) in three dimensions that is described
by the single-channel model Hamiltonian \cite{chevy2006,combescot2007,Liu2005PRA},
\begin{alignat}{1}
H=  \sum_{\mathbf{k}}\left[\left(\epsilon_{\mathbf{k}}-\mu\right)c_{\mathbf{k}\uparrow}^{\dagger}c_{\mathbf{k}\uparrow}+\left(\epsilon_{\mathbf{k}}-\mu_{\downarrow}\right)c_{\mathbf{k}\downarrow}^{\dagger}c_{\mathbf{k}\downarrow}\right] 
  +\frac{U}{V}\sum_{\mathbf{q},\mathbf{k},\mathbf{k}'}c_{\mathbf{k}\uparrow}^{\dagger}c_{\mathbf{q}-\mathbf{k}\downarrow}^{\dagger}c_{\mathbf{q}-\mathbf{k}'\downarrow}c_{\mathbf{k}'\uparrow},\label{eq:hami}
\end{alignat}
where $\epsilon_{\mathbf{k}}\equiv\hbar^{2}\mathbf{k}^{2}/(2m)$,
$\mu_{\uparrow}=\mu$ and $\mu_{\downarrow}$ are the chemical potentials
of spin-up and spin-down atoms, respectively, for atoms of mass $m$,
and $U<0$ is the bare attractive interatomic interaction strength.
We renormalize the interaction in the usual prescription in terms
of the $s$-wave scattering length $a$, 
\begin{equation}
\frac{1}{U}=\frac{m}{4\pi\hbar^{2}a}-\sum_{\mathbf{k}}\frac{m}{\hbar^{2}\mathbf{k}^{2}}.\label{eq:renormalization}
\end{equation}

\subsection{Many-body $T$-matrix theory}

We start the calculation of the impurity thermal Green's function,
\begin{alignat}{1}
G_{\downarrow}\left(\mathbf{k},i\omega_{m}\right)=\frac{1}{i\omega_{m}-\left(\epsilon_{\mathbf{k}}-\mu_{\downarrow}\right)-\Sigma_{\downarrow}\left(\mathbf{k},i\omega_{m}\right)},\label{eq: gfdown}
\end{alignat}
from the many-body $T$-matrix approximation, where the fermionic
Matsubara frequencies are given by $\omega_{m}\equiv(2m+1)\pi/\beta$
for the inverse temperature $\beta=1/(k_{B}T)$ and any integer $m$.
We take spin-down atoms as the impurities and assume that in the
limit of large polarization, $x=n_{\downarrow}/n\ll1$, the majority
spin-up atoms are not affected by the interactions to a leading order
approximation\footnote{We have checked this approximation by numerically calculating the
majority Green's function at finite impurity and temperature, however to the higher order this approximation may break down \cite{Tajima2018}.}. Taking the majority Green's functions as the non-interacting Green's
function, 
\begin{equation}
G_{\uparrow}^{(0)}\left(\mathbf{k},i\omega_{m}\right)=\frac{1}{i\omega_{m}-\left(\epsilon_{\mathbf{k}}-\mu\right)},\label{eq: gf0up}
\end{equation}
the chemical potential of the majority atoms $\mu$ as a function
of temperature can then be found as $\mu^{(0)}(T)\overset{T\rightarrow0}{\simeq}\varepsilon_{\textrm{F}}=\hbar^{2}(6\pi^{2}n)^{2/3}/(2M)$.
Using the many-body T-matrix formalism set out in Appendix~\ref{app:tmatrix}, we find the analytically continued impurity Green's function 
\begin{alignat}{1}
G_{\downarrow}(\mathbf{k},\omega^{+})=\frac{1}{\omega^{+}-\left(\epsilon_{\mathbf{k}}-\mu_{\downarrow}\right)-\Sigma_{\downarrow}(\mathbf{k},\omega^{+})},
\end{alignat}
and self energy $\Sigma_{\downarrow}(\mathbf{k},\omega^{+})$. 
The spectral function is found from the analytically continued Green's
function $A(\mathbf{k},\omega^{+})=-2{\rm Im}\,G(\mathbf{k},\omega^{+})$,
which is given by 
\begin{alignat}{1}
 & A_{\downarrow}(\mathbf{k},\omega^{+})\nonumber \\
 & =\frac{{\rm Im}\,\Sigma_{\downarrow}(\mathbf{k},\omega^{+})}{(\omega^{+}-\xi_{\mathbf{k}}-{\rm Re}\,\Sigma_{\downarrow}(\mathbf{k},\omega^{+}))^{2}+({\rm Im}\,\Sigma_{\downarrow}(\mathbf{k},\omega^{+}))^{2}}.\label{eq:spectral}
\end{alignat}
The impurity chemical potential is computed self-consistently such that we find the
density, 
\begin{alignat}{1}
n_{\downarrow}=\int\frac{d\mathbf{k}}{(2\pi)^{3}}\int_{-\infty}^{\infty}\frac{d\omega}{2\pi}A_{\downarrow}(\mathbf{k},\omega)f(\omega).
\end{alignat}

\subsection{Virial expansion theory}

At high temperature, the virial expansion is a powerful tool to understand
strongly-correlated many-body systems \cite{Liu2009,Kaplan2011,Leyronas2011,Liu2013}, 
and has been found to be successful in describing the high-temperature
properties of ultracold gases \cite{Nascimbene2010,Ku2012}.
It is an expansion in terms of the small fugacity $z_{\sigma}=e^{\beta\mu_{\sigma}}$
for each component $\sigma$. The virial expansion of the single-particle
Green function in a spin-balanced Fermi gas was recently developed
by Sun and Leyronas \cite{sun2015} and the technique was applied
to investigate the Bose polaron at high temperature \cite{sun2017prl,sun2017pra}.
In this work, we apply the same technique to understand the Fermi
polaron at high temperature. As the theory was already well-documented
by Sun \textit{et al.} \cite{sun2015,sun2017prl,sun2017pra}, here
we only give a brief introduction on the essential idea and present
the details in Appendix~\ref{app:Green}. Following the work of Refs.~\cite{sun2017prl,sun2017pra}
we expand the impurity Green's function, taking only the diagrams
that do not contain any orders of the impurity fugacity, as these
diagrams will contribute at a higher order.

The expansion starts from the non-interacting Green's function in
momentum space and imaginary time: 
\begin{alignat}{1}
G_{\sigma}^{(0)}(\mathbf{k},\tau)=e^{-(\xi_{\mathbf{k}}-\mu_{\sigma})\tau}\left[-\Theta(\tau)+n_{{\rm F}}(\xi_{\mathbf{k}}-\mu_{\sigma})\right],\label{eq:Green_nonint}
\end{alignat}
where $\Theta(\tau)$ is the Heaviside step function and $n_{{\rm F}}(x)=1/\left(e^{\beta x}+1\right)$
is the Fermi distribution. We expand the Fermi distribution in powers
of the fugacity: 
\begin{alignat}{1}
G_{\sigma}^{(0)}(\mathbf{k},\tau)=e^{\mu_{\sigma}\tau}\left[\sum_{n\ge0}G^{(0,n)}(\mathbf{k},\tau)z_{\sigma}^{n}\right],
\end{alignat}
where we define 
\begin{alignat}{1}
G^{(0,0)}(\mathbf{k},\tau) & =-\Theta(\tau)e^{-\xi_{\mathbf{k}\tau}},\\
G^{(0,n)}(\mathbf{k},\tau) & =(-1)^{n-1}e^{-\xi_{\mathbf{k}\tau}}e^{-n\beta\xi_{\mathbf{k}}}.
\end{alignat}
First, we note the dependence on the chemical potential $\mu_{\sigma}$
does not enter into our definition of $G^{(0,0)}$ and $G^{(0,n)}$,
the chemical potential is found in the fugacity, $z_{\sigma}^{n}$,
and the global $e^{\mu_{\sigma}\tau}$ factor. Because of the step
function the lowest order term $G^{(0,0)}$ is retarded, and the $G^{(0,n)}$
is not retarded. When we expand the full Green's function diagrammatically,
we will in general expand free propagators that run forward, as well
as some that run backward in imaginary time. A particle running backwards
in imaginary time that comes in from the medium and scattering with
a particle is a hole scattering process. Any diagram which involves
particle-particle scattering, lines which run forward in time, we
will be able to write in terms of bare $G^{(0,0)}$ at the lowest
order in the fugacity. A diagram which then contains a backward running
line (a hole) can then be expanded in terms of $G^{(0,n)}z_{\sigma}^{n}$,
and is represented by a line with $n$ slashes. 

We show in Appendix~\ref{app:thermo}
the derivation of the thermodynamic potential to the third order of
the fugacity and in Appendix~\ref{app:Green} the expansion of the
Green's function in orders of the fugacity and the detailed diagrammatic construction for the impurity self-energy
(i.e., Figs.~\ref{fig:sig_2} and \ref{fig:sig_3}). We use the rf spectra calculated from the virial
expansion to compare with the $T$-matrix spectra in the high temperature
regime.

\subsection{Quasi-particle properties}

Once the impurity Green's function and chemical potential have been
determined, we directly calculate the quasi-particle properties of
the polaron. Near a quasi-particle excitation, we may separate the
analytically continued Green's function into two parts: a pole contribution
(from quasi-particle) plus an incoherent background, 
\begin{equation}
G_{\downarrow}=\frac{\mathcal{Z}}{\omega-\hbar^{2}\mathbf{k}^{2}/(2m^{*})+\mu_{\downarrow}-E_{P}+i\gamma/2}+\cdots,\label{eq:gfRdown}
\end{equation}
where $\mathcal{Z}$ is the quasi-particle residue, $E_{P}$ is the
energy of the polaron with effective mass $m^{*}$, and $\gamma$
is the decay rate of the polaron. This gives rise to a polaron spectral
function $A_{\downarrow}(\mathbf{k},\omega)=-2\textrm{Im}G_{\downarrow}(\mathbf{k},\omega)$
\cite{baarsma2012,hui2017}, 
\begin{equation}
A_{\downarrow}\left(\mathbf{k},\omega\right)=2\pi\mathcal{Z}\delta\left(\omega+\mu_{\downarrow}-\frac{\hbar^{2}\mathbf{k}^{2}}{2m^{*}}-E_{P}\right)+\cdots.\label{eq:Adown}
\end{equation}
The attractive and repulsive polarons are found from the poles of
the Green's function, $E_{P}=\omega_{{\rm pole}}+\mu_{\downarrow}$
\cite{combescot2007}, where the polaron energy is related to the
self-energy by, 
\begin{alignat}{1}
E_{P}= & \textrm{Re}\Sigma_{\downarrow}\left(\mathbf{k}=0,E_{P}-\mu_{\downarrow}\right).\label{eq:polar}
\end{alignat}
The spectral weight, effective mass, and decay rate are given by 
\begin{eqnarray}
\mathcal{Z} & = & \left.\left(1-\frac{\partial\mathbf{\textrm{Re}}\Sigma_{\downarrow}}{\partial\omega}\right)^{-1}\right|_{\mathbf{k}=0,\omega=\omega_{{\rm pole}}},\label{eq:Residue}\\
\frac{m}{m^{*}} & = & \left.\left(1+\frac{\partial\mathbf{\textrm{Re}}\Sigma_{\downarrow}}{\partial\epsilon_{\mathbf{k}}}\right)\left(1-\frac{\partial\mathbf{\textrm{Re}}\Sigma_{\downarrow}}{\partial\omega}\right)^{-1}\right|_{\mathbf{k}=0,\omega=\omega_{{\rm pole}}},\label{eq:inverseMass}\\
\gamma & = & -2\mathcal{Z}\textrm{Im}\Sigma_{\downarrow}(0,\omega_{{\rm pole}}).
\end{eqnarray}

\begin{figure}
\centering{}\includegraphics[width=0.45\textwidth]{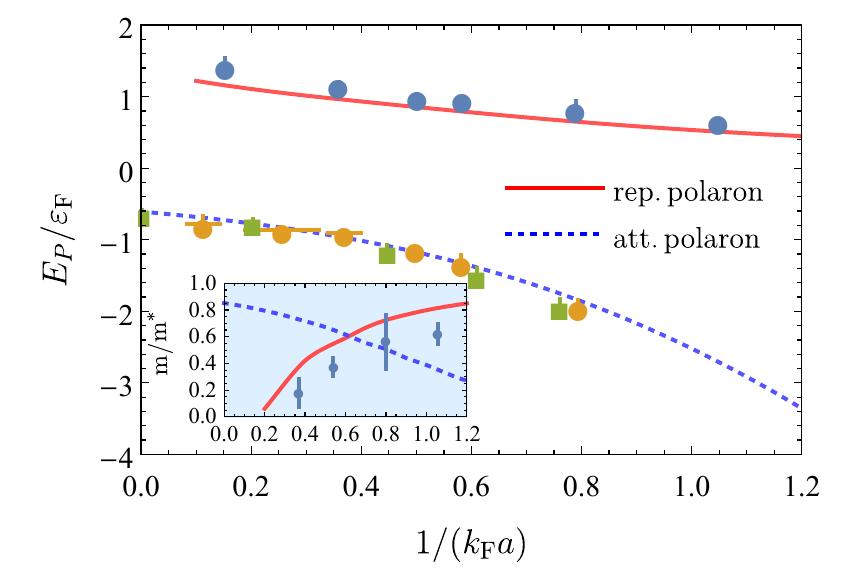}
\caption{ The attractive (lower branch) and repulsive (upper branch) polaron
energies as a function of the interaction strength for an impurity
density $x\equiv n_{\downarrow}/n=0.01$ and temperature $T=0.03T_{{\rm F}}$.
We show for comparison the experimental results from Ref.~\cite{Scazza2017}
(circular symbols) and the results from Ref.~\cite{Schiro2009} (square
symbols). The inset shows the inverse effective mass of the repulsive
polaron (solid red), attractive polaron (blue dashed), and experimental
results Ref.~\cite{Scazza2017} (circular symbols). \label{fig:zerotemppol}}
\end{figure}

As an example of the quasi-particle properties of the analytically
continued Green's function we show the attractive and repulsive polaron
energies in Fig.~\ref{fig:zerotemppol} found from Eq.~\eqref{eq:polar}
using the many-body $T$-matrix theory. Fixing the temperature to
effectively zero, $T=0.03T_{{\rm F}}(\equiv\varepsilon_{\textrm{F}}/k_{B})$,
and the impurity density to $x=0.01$ we show the attractive (lower
branch) and repulsive (upper branch) polaron energies as a function
of the dimensionless interaction strength $(k_{\textrm{F}}a)^{-1}$,
where $k_{\textrm{F}}=(3\pi^{2}n)^{1/3}$. We find a reasonable agreement
with the experiments of Ref.~\cite{Schiro2009} (square symbols)
and Ref.~\cite{Scazza2017} (circular symbols), as has been found
in previous works, for both the attractive and repulsive polarons \cite{Massignan2011,Tajima2018}.
For the repulsive branch, our calculations do not attempt to find
the polaron energy below a threshold interaction strength of $(k_{{\rm F}}a)^{-1}=0.1$
\cite{schmidt2011,Tajima2018}. The inset shows the effective mass for the attractive
(blue dashed) and repulsive (red solid) polarons. We see that for
the repulsive polaron the effective mass is consistently over-estimated
for all interaction strengths compared to the experimental results.
The over-estimation in the predicted effective mass at zero temperature
was noted earlier in Ref. \cite{Scazza2017}.

\subsection{Radio-frequency spectroscopy}

The single-particle properties of the polaron can be experimentally
probed with radio-frequency (rf) spectroscopy \cite{Schunck2008,Stewart2008},
which has been used in experiments to find the energy, effective mass,
and residue of the attractive and repulsive polarons \cite{Schiro2009,Scazza2017,kohl2012,kohstall2012}.
Theoretically, by calculating the rf spectra we can connect our many-body
$T$-matrix results of the quasi-particle properties of the polaron
at finite temperature and impurity density directly to the experimentally
observed rf spectra. With our knowledge of the spectral function in
Eq.~\eqref{eq:spectral}, we calculate the \textit{direct} rf and
\textit{reverse} rf spectroscopy of the polaron and find the position
of the maximum of the peaks, which corresponds to the attractive and
repulsive polaron energies, and the full width half maximum, which
can be viewed as the lifetime of the polaron.

We consider a three-component Fermi gas with the majority $\left|\uparrow\right\rangle $
component and minority components $\left|i\right\rangle $ and $\left|f\right\rangle $.
Within the linear response frame work the transition from an initial
to final state is \cite{punk2007,massignan2008,haussmann2009,chen2009,Zwerger09,torma2016},
\begin{alignat}{1}
I(\omega_{{\rm rf}})=2\Omega^{2}{\rm Im}\,\chi(\mathbf{k}=0,\mu_{f}-\mu_{i}-\omega_{{\rm rf}}),
\end{alignat}
where $\Omega$ is the Rabi frequency, $\mu_{i}$ and $\mu_{f}$ are
the initial and final state chemical potentials, and $\omega_{{\rm rf}}$
is the rf frequency. At finite temperature the retarded correlation
function is found from the time-ordered correlation function, 
\begin{alignat}{1}
\chi(\mathbf{r},\mathbf{r}',\tau)=\langle T_{\tau}\psi_{f}^{\dagger}(\mathbf{r},\tau)\psi_{i}(\mathbf{r},\tau)\psi_{f}(\mathbf{r},0)\psi_{i}^{\dagger}(\mathbf{r},0)\rangle.
\end{alignat}
The calculation of the analytically continued correlation function
contains several different diagrammatic contributions \cite{chen2009,schmidt2011},
for our calculation we will assume there are no final state interactions
and ignore the higher order vertex corrections \cite{perali2008,Pieri2009}.
The spectral response function in the domain of Matsubara frequencies
then becomes 
\begin{alignat}{1}
\chi(i\nu_{n})=\frac{1}{\beta}\sum_{i\omega_{m}}\int\frac{d\mathbf{k}}{(2\pi)^{3}}G_{f}(\mathbf{k},i\omega_{m})G_{i}(\mathbf{k},i\nu_{n}+i\omega_{m})
\end{alignat}
and the rf response on the real axis is 
\begin{alignat}{1}
I(\omega)= & \Omega^{2}\int\frac{d\mathbf{k}}{(2\pi)^{3}}\int\frac{d\epsilon}{2\pi}
 f(\epsilon)A_{f}(\mathbf{k},\epsilon+\omega+\mu_{i}-\mu_{f})A_{i}(\mathbf{k},\epsilon).
\end{alignat}
Here, $A_{\sigma}(\mathbf{q},\omega)$ corresponds to the spectral
function of the state $\sigma$. In the calculation of the rf spectra
we ignore the occupation of the final state, however at finite temperature
the occupation is, in principle, non-zero. Taking the final state
to then become $A_{f}(\mathbf{q},\omega)=2\pi\delta(\omega-\epsilon_{\mathbf{q}}+\mu_{f})$
and the initial state to be the minority the \textit{direct} rf spectroscopy
is given by 
\begin{alignat}{1}
I(\omega)=-\Omega^{2}\int\frac{d\mathbf{k}}{(2\pi)^{3}}f(\epsilon_{\mathbf{k}}-\mu_{\downarrow}-\omega)A_{\downarrow}(\mathbf{k},\epsilon_{\mathbf{k}}-\mu_{\downarrow}-\omega).
\end{alignat}
The \textit{reverse} rf spectroscopy is given by flipping the spins
from an initial non-interacting state to a final state which is strongly
interacting, i.e. the unoccupied minority state, 
\begin{alignat}{1}
I(\omega)=-\Omega^{2}\int\frac{d\mathbf{k}}{(2\pi)^{3}}f(\epsilon_{\mathbf{k}}-\mu_{i})A_{\downarrow}(\mathbf{k},\epsilon_{\mathbf{k}}-\mu_{\downarrow}+\omega).
\end{alignat}
where $\mu_{i}$ can be determined from the non-interacting impurity
in state $\left|i\right\rangle $ and the final chemical potential
in the spin-flipped state $\left|f\right\rangle $ is $\mu_{\downarrow}$.
As a check to our calculation of the direct rf spectra we can calculate the
number density, i.e. 
\begin{alignat}{1}
n_{\downarrow}=\int\frac{d\omega}{2\pi}I(\omega),
\end{alignat}
which holds if we set the Rabi frequency $\Omega=1$.

\begin{figure}
\centering{}\includegraphics[width=0.9\textwidth]{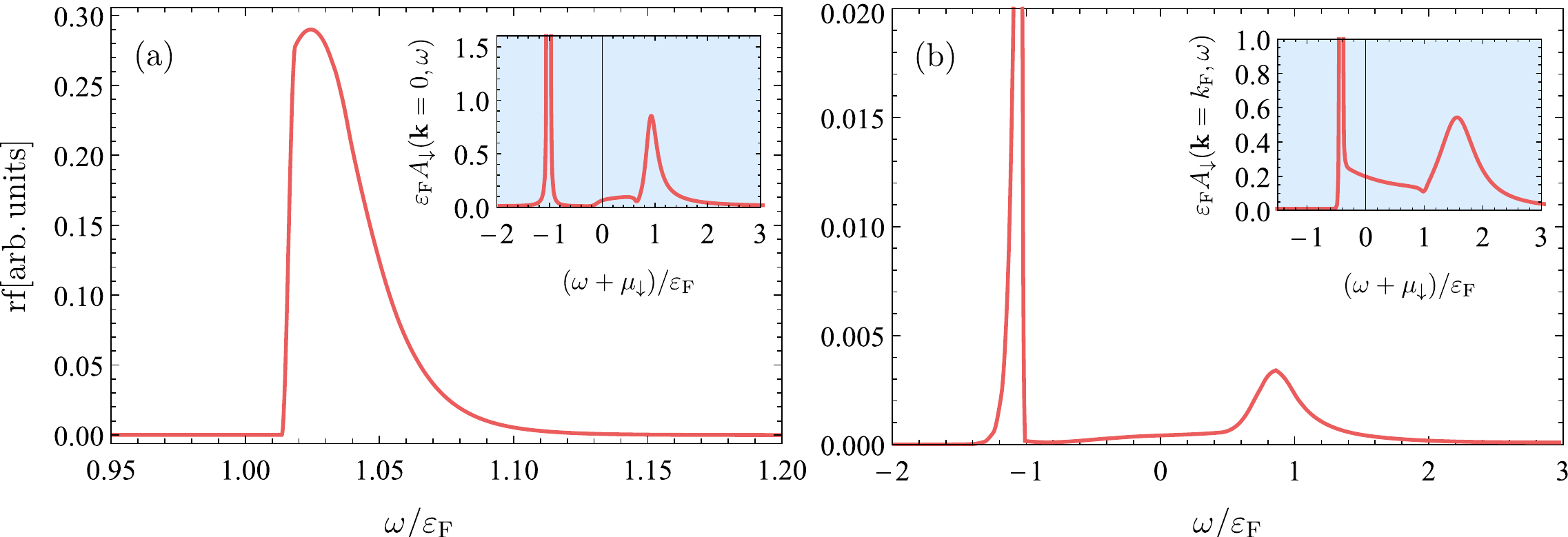} \caption{(a) The direct rf spectroscopy in arbitrary units for an interaction
strength of $1/(k_{{\rm F}}a)=0.4$, an impurity concentration $x=0.01$
and temperature $T=0.03T_{{\rm F}}$. (b) The reverse rf spectroscopy
at the same condition. The two insets plot the corresponding dimensionless
spectral function at zero momentum, $A_{\downarrow}(\mathbf{k}=0,\omega)$,
and at the Fermi momentum, $A_{\downarrow}(\mathbf{k}=k_{{\rm F}},\omega)$.
\label{fig:zerotemp_spec}}
\end{figure}

As an example of the two rf spectroscopy schemes we plot in Fig.~\ref{fig:zerotemp_spec}(a)
the direct rf spectra and in Fig.~\ref{fig:zerotemp_spec}(b) the
reverse rf spectra for an interaction strength of $1/(k_{{\rm F}}a)=0.4$,
impurity concentration $x=0.01$, and temperature $T=0.03T_{{\rm F}}$,
calculated by using the many-body $T$-matrix theory. The peak value
in the direct spectrum corresponds to the attractive polaron energy
(more precisely, $-E_{P}$). For the reverse rf spectroscopy the repulsive
polaron is found from the positive peak and the peak at negative frequencies
is the attractive polaron energy. We see in both spectra the attractive
polaron peak is asymmetric and for the reverse rf scheme the peak
at positive frequency is significantly broader, indicating the finite
lifetime of the repulsive polaron.  The polaron energies are the same as those 
found in Fig.~\ref{fig:zerotemppol} with the quasi-particle description, as expected.

The two insets show the spectral
function calculated at zero momentum and the Fermi momentum, respectively,
and we see how the spectral weight shifts from the negative peak to
the positive energy peak as the momentum increases. Going from $k=0$ to $k=k_{\rm F}$, both peaks
shift up in energy by an amount comparable to the polarons' kinetic
energies, $(m/m^{*})E_{\rm F}$, where $(m/m^{*})$ for the attractive and repulsive
branches may be read out from the inset of Fig.~\ref{fig:zerotemppol} ($\simeq0.7$ for the
attractive polaron and $\simeq0.4$ for the repulsive polaron). For the spectral
functions and the reverse spectra in Fig.~\ref{fig:zerotemp_spec}
(b) we have added a finite width with a small imaginary part to make
the sharp peaks of the attractive polaron visible.

At zero temperature and for a single impurity we expect the attractive
polaron energy from the rf spectroscopy to be a sharp $\delta$-function
peak; the rf pulse provides the energy required to excite
the polaron into the final state. As the temperature and impurity
density increases, the width of the rf spectra increases due to the
finite number of states which are now occupied in the spectral function
of the imbalanced gas. Within the $T$-matrix scheme considered here
we need a finite impurity population and the lowest temperature we
can consider is $T=0.03T_{{\rm F}}$, and we expect the spectra to
have a finite asymmetrical width which shifts the energy of the polaron.
As the temperature and impurity increases we expect the widths and
peak positions of the two rf spectroscopy schemes to change by differing
amounts, i.e. we do not expect the temperature and impurity dependence
to be the same.

\section{Fermi polaron near Fermi degeneracy at unitarity}

\label{sec:hightemp}

\begin{figure}
\centering{}\includegraphics[width=0.9\textwidth]{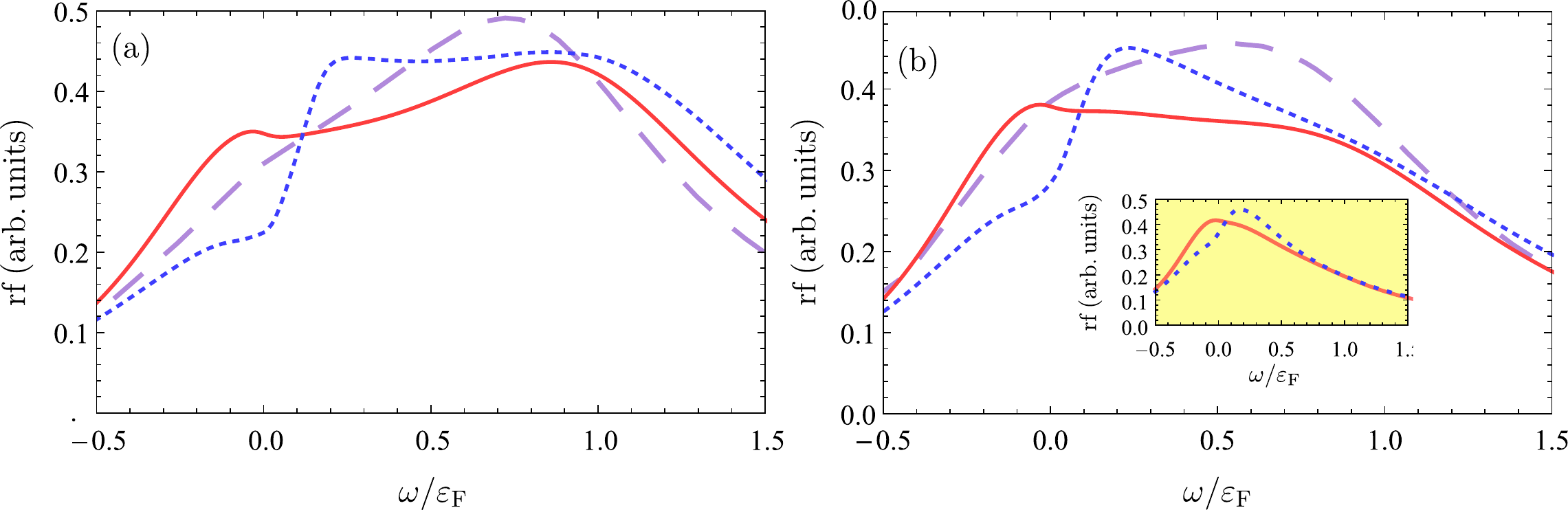}
\caption{Direct rf spectroscopy from the second (red solid), third (blue dotted)
virial expansion, and $T$-matrix (purple dashed) at the interaction
strength $1/(k_{{\rm F}}a)=0$ for temperatures in (a) $T=1.25T_{{\rm F}}$
and (b) $T=1.5T_{{\rm F}}$. Insert: rf spectroscopy comparing the
second and third virial expansion at the temperature $T=2.0T_{{\rm F}}$.}
\label{fig:rf3rd} 
\end{figure}

In the high temperature regime we expect the quasi-particle description
of the polaron to break down and determining this transition temperature
is non-trivial. Motivated by recent experiments at MIT \cite{Struck2017},
we explore the breakdown of the attractive polaron in the unitary
limit as a function of temperature using the spectral function calculated
from the $T$-matrix theory. We calculate the attractive polaron energy
from the peak value of the rf spectra and the lifetime of the quasi-particle
excitation from the FWHM \cite{Schiro2009,Struck2017}.


\begin{figure}
\centering{}\includegraphics[width=0.45\textwidth]{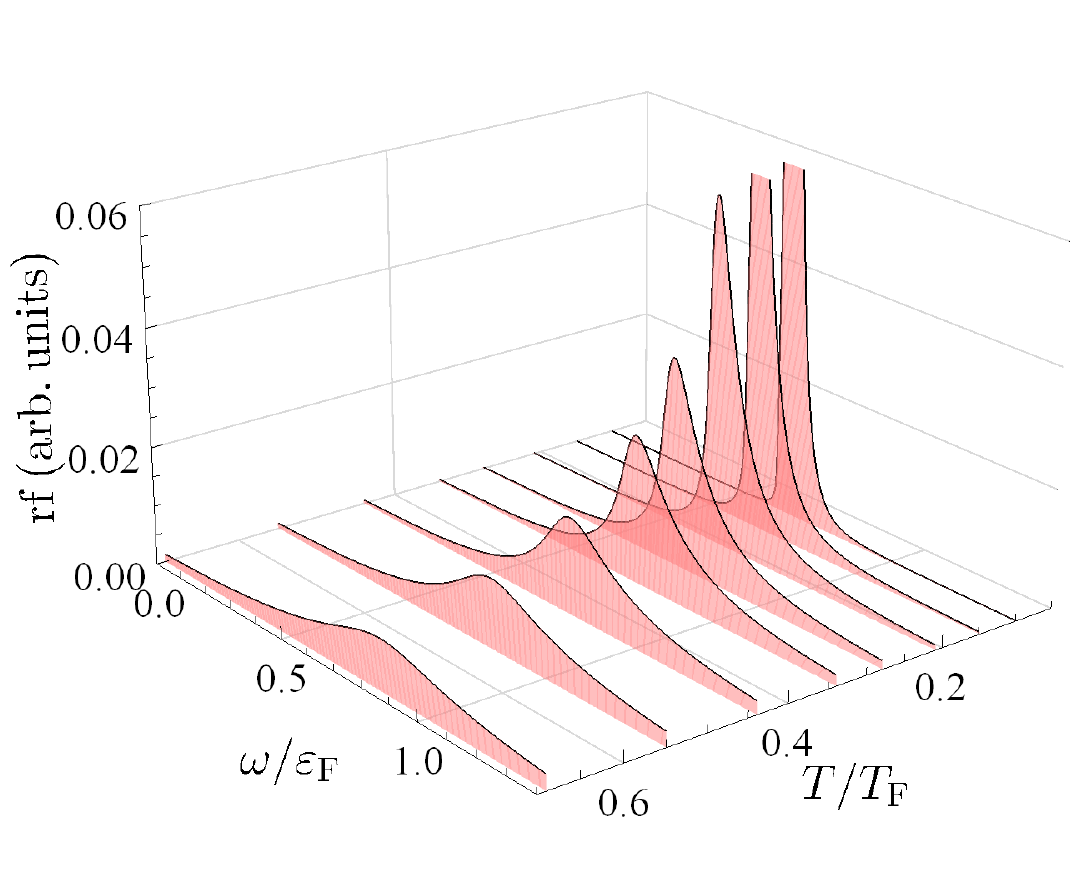}
\caption{Direct rf spectroscopy in arbitrary units from the $T$-matrix Green's
function at $1/(k_{{\rm F}}a)=0$ for temperatures from $T=0.1T_{{\rm F}}$
to $T=0.7T_{{\rm F}}$ as a function of rf pulse $\omega/\varepsilon_{{\rm F}}$.}
\label{fig:rfhigh} 
\end{figure}

In Fig.~\ref{fig:rf3rd} we plot the direct rf spectroscopy from
the $T$-matrix (purple dashed) spectral function and the high temperature
virial expansion calculated at the second (red solid) and third order
(blue dotted) at two temperatures (a) $T=1.25T_{{\rm F}}$ and (b)
$T=1.5T_{{\rm F}}$, for the interaction strength $1/(k_{{\rm F}}a)=0$
and the finite impurity density $x=0.1$ \cite{Struck2017}. We see
in Fig.~\ref{fig:rf3rd}(a) that for all three spectra there is a
peak at finite rf frequency suggesting the formation of a quasi-particle
with finite energy. However, the peaks are very broad and asymmetrical,
indicating that a quasi-particle is not well defined. There are additional
structures in the virial spectra, i.e., the large asymmetry between
the peaks and the sharp increase in the third order at $\omega\simeq0$.
We attribute them to the finite number of terms in the expansion of
the virial Green's function; in the strongly interacting regime physics
of more than three-body contributions will play a significant role.
As we go to higher temperature, $T=1.5T_{{\rm F}}$ in Fig.~\ref{fig:rf3rd}(b)
the peak at finite rf frequency is transferred to the peak at $\omega\simeq0$
for all three spectra, and the system is becoming weakly interacting.
The inset in Fig.~\ref{fig:rf3rd} shows the second and third virial
spectra at the temperature $T=2T_{{\rm F}}$. The two peaks have merged
into a single peak at $\omega\simeq0$, indicating that there is no
longer any quasi-particle in the system and it is weakly interacting.

For temperatures below the Fermi temperature, $T=T_{{\rm F}}$, we
expect the virial and $T$-matrix spectra will qualitatively give
the same behavior, however, the virial expansion is expected to break
down as the temperature is lowered and the fugacity becomes larger.
There is no definite method to define a threshold temperature at which
the expansion has broken down. We see in the calculation of the Green's
function at the third order, the spectral function becomes unphysical
for values of fugacity $z_{\downarrow}\geq0.3$ ($T\simeq1.25T_{{\rm F}}$),
where the spectral function becomes negative for some values of $\omega$
at small momenta. We expect that this will be canceled off by higher
order terms in the virial expansion, and we take this temperature
as a lower-bound for the validity of the virial expansion.

\begin{figure}
\centering{}\includegraphics[width=0.45\textwidth]{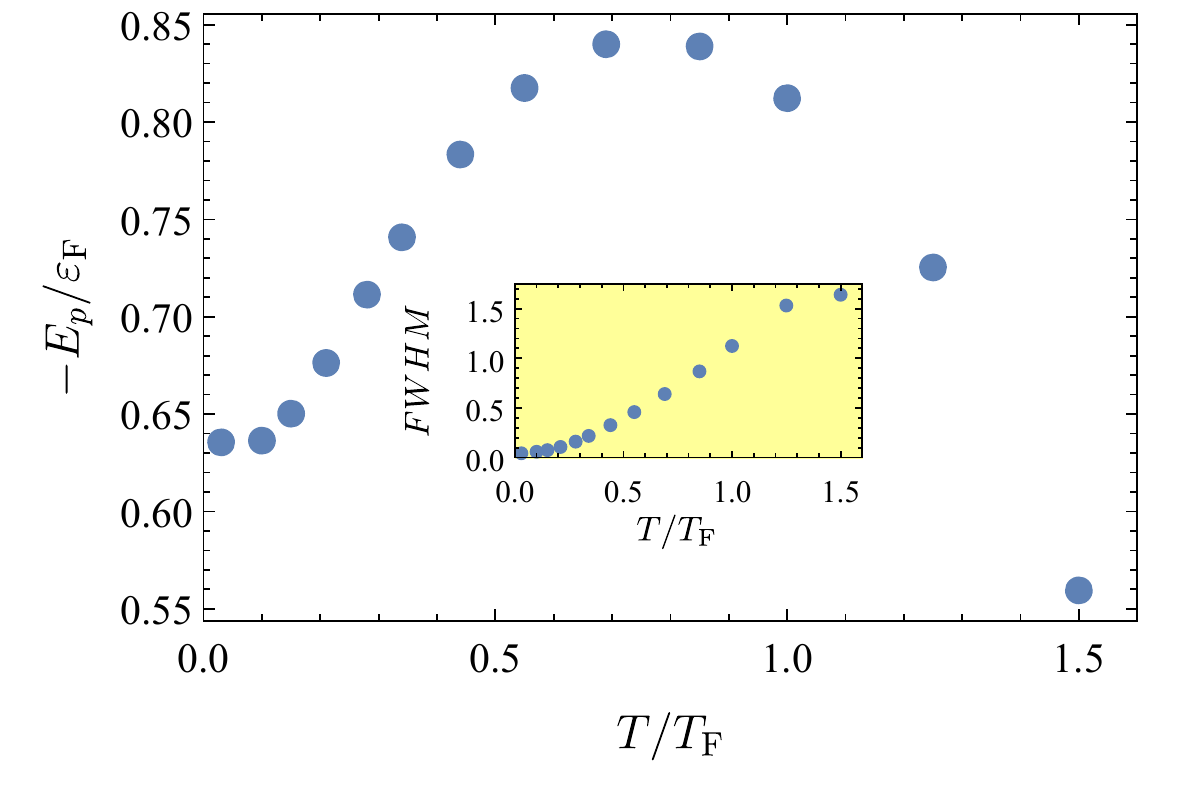} \caption{The temperature dependence of the attractive polaron energy at unitarity
for the impurity concentration $x=0.1$ from the peak position of
the rf spectra. The inset shows the full width half maximum in units
of $\varepsilon_{F}$. \label{fig:EPFWHM}}
\end{figure}

Moving to the low temperature regime, in Fig.~\ref{fig:rfhigh} we
show the temperature evolution of the rf spectra calculated from the
$T$-matrix theory at temperatures from $T=0.1T_{{\rm F}}$ to $T=0.7T_{{\rm F}}$
and at the impurity concentration of $x=0.1$ \cite{Struck2017}.
We clearly see that there remains a definite peak and the spectra broaden
as the temperature increases. We extract the peak values of the spectra
and the FWHM and plot them in Fig.~\ref{fig:EPFWHM}. We find that
the polaron energy $-E_{P}$ increases and has a maximum around $T\simeq0.8T_{{\rm F}}$ 
as the attraction between polaron and the medium increases. The energy of the polaron then decreases 
for temperatures above $T\simeq0.8T_{{\rm F}}$ as the the FWHM becomes on the order of the Fermi energy and the Fermi surface broadens.
For low temperatures the FWHM is increasing approximately
as $(T/T_{{\rm F}})^{2}$ and becomes greater than the polaron energy at
temperatures $T>0.8T_{{\rm F}}$. 
We propose that the quasi-particles
are well defined if their inverse lifetime is less than their excitation
energy. Thus, we conclude that the attractive polaron in the unitary
limit remains well defined up to temperatures of about 
$T\simeq0.8T_{{\rm F}}$\footnote{A non-zero impurity concentration may 
also break down the quasi-particle
picture, in this work we have chosen a finite density of $x=0.1$
as a realistic choice for an experimental setup \cite{Struck2017}.}.

\section{Conclusion}

\label{sec:conc}

In summary, using the many-body $T$-matrix approximation
and high temperature virial expansion, we have calculated the direct
rf spectroscopy for a range of temperatures at unitarity and discussed
the breakdown of the quasi-particle description of the attractive
polaron as temperature increases. In the high temperature regime,
where the virial expansion is valid, we found qualitative agreement
between the rf spectra obtained from the $T$-matrix scheme and virial
expansion, showing the failure of the quasi-particle description.
In the low temperature regime, where the $T$-matrix theory is reliable,
we have calculated the FWHM of the rf spectra and have found that
the quasi-particle description is well defined for temperatures below
$T\simeq0.8T_{{\rm F}}$, where the FWHM becomes smaller than the
absolute value of the polaron energy.

\section*{Acknowledgment}
We thank Jia Wang for reading of the manuscript and Zhenjie Yan for
their comments. Our research was supported by Australian Research
Council's (ARC) Discovery Projects: DP140100637, FT140100003 and DP180102018
(XJL), FT130100815 and DP170104008 (HH). 

Note added: Recently, the  high temperature behavior of the polaron was examined experimentally by Zhenjie {\it et al.} \cite{zhenjie2018}. We note that similar results were obtained for the breakdown of the polaron description at unitary. However, we note that for temperatures above $T\simeq0.75T_{\rm F}$ the authors find a sharp jump in the position of the global maxium to $\omega\simeq0$, which is not captured by our $T$-matrix approximation.

\appendix

\section{Many-body T-matrix} \label{app:tmatrix}

To use the well-established many-body $T$-matrix theory to find the
impurity Green's function $G_{\downarrow}\left(\mathbf{k},i\omega_{m}\right)$
\cite{Haussmann93,Liu2005PRA,Combescot2006,Ohashi2009}, we sum all
of the ladder-type diagrams, and obtain the self-energy, 
\begin{equation}
\Sigma_{\downarrow}=k_{B}T\sum_{\mathbf{q},i\nu_{n}}G_{\uparrow}^{(0)}\left(\mathbf{q}-\mathbf{k},i\nu_{n}-i\omega_{m}\right)\Gamma\left(\mathbf{q},i\nu_{n}\right),\label{eq:selfenergy1}
\end{equation}
where the vertex function $\Gamma$ can be written through the Bethe-Salpeter
equations, 
\begin{equation}
\Gamma\left(\mathbf{q},i\nu_{n}\right)=\frac{1}{U^{-1}+\chi\left(\mathbf{q},i\nu_{n}\right)},\label{eq:vertexfunction}
\end{equation}
with the pair propagator $\chi(\mathbf{q},i\nu_{n})$, 
\begin{equation}
\chi=k_{B}T\sum_{\mathbf{k},i\omega_{m}}G_{\uparrow}^{(0)}\left(\mathbf{q}-\mathbf{k},i\nu_{n}-i\omega_{m}\right)G_{\downarrow}^{(0)}\left(\mathbf{k},i\omega_{m}\right),\label{Eq:propagator2p}
\end{equation}
and the bosonic Matsubara frequencies, $\nu_{n}\equiv2n\pi/\beta$,
for integer $n$.

The closed set of equations, \eqref{eq: gfdown} to \eqref{Eq:propagator2p},
can be solved directly with Matsubara frequencies \cite{hui2017}.
Alternatively, we can analytically continue the Matsubara frequencies
to the real axis. This will allow us to calculate the spectral function
without numerically continuing to real frequencies \cite{Veillette2008}.
The analytically continued impurity Green's function is given by 
\begin{alignat}{1}
G_{\downarrow}(\mathbf{k},\omega^{+})=\frac{1}{\omega^{+}-\left(\epsilon_{\mathbf{k}}-\mu_{\downarrow}\right)-\Sigma_{\downarrow}(\mathbf{k},\omega^{+})},
\end{alignat}
where $\omega^{+}\equiv\omega+i0^{+}$ and the self-energy function
now takes the form \cite{Rohe2001}, 
\begin{alignat}{1}
\Sigma_{\downarrow} & (\mathbf{k},\omega^{+})=\nonumber \\
 & \int\frac{d\mathbf{q}}{(2\pi)^{3}}\frac{d\epsilon}{\pi}\biggl[b(\epsilon)G_{\uparrow}^{(0)}(\mathbf{k}-\mathbf{q},\epsilon-\omega^{+}){\rm Im}\Gamma(\mathbf{q},\epsilon^{+})\nonumber \\
 & -f(\epsilon){\rm Im}G_{\uparrow}^{(0)}(\mathbf{k},\epsilon^{+})\Gamma(\mathbf{k}+\mathbf{q},\epsilon+\omega^{+})\biggl],
\end{alignat}
where $f(z)=(\exp(\beta z)+1)^{-1}$ and $b(z)=(\exp(\beta z)-1)^{-1}$
are the Fermi and Bose distributions respectively \footnote{It should be noted that there is an additional contribution to the
self-energy from the bound state when there is a pole in the vertex
function.}. Performing the Matsubara sum analytically the vertex function is
given by 
\begin{alignat}{1}
\Gamma^{-1} & \left(\mathbf{q},\omega^{+}\right)\nonumber \\
 & =\frac{m}{4\pi a}-\sum_{\mathbf{k}}\left[\frac{1-f(\xi_{\mathbf{k}+\mathbf{q}/2}^{\uparrow})-f(\xi_{\mathbf{k}-\mathbf{q}/2}^{\downarrow})}{\omega^{+}-\xi_{\mathbf{k}-\mathbf{q}/2}^{\uparrow}-\xi_{\mathbf{k}+\mathbf{q}/2}^{\downarrow}}+\frac{1}{2\epsilon_{\mathbf{k}}}\right],
\end{alignat}
where $\xi_{\mathbf{k}}^{\sigma}=\epsilon_{\mathbf{k}}-\mu_{\sigma}$.
We then find the imaginary part of the analytically continued self-energy,
\begin{alignat}{1}
{\rm Im}\,\Sigma_{\downarrow}(\mathbf{k},\omega)= & \int\frac{d^{3}q}{(2\pi)^{3}}\frac{d\epsilon}{2\pi}\left(b(\epsilon)+f(\epsilon-\omega)\right)\nonumber \\
 & \times{\rm Im}\Gamma(\mathbf{q},\epsilon)\,{\rm Im}G_{\uparrow}^{(0)}(\mathbf{q}-\mathbf{k},\epsilon-\omega),
\end{alignat}
and we calculate the real part of the self-energy from the Kramers-Kronig
relation, 
\begin{alignat}{1}
{\rm Re}\,\left[f(z)\right]=\frac{1}{\pi}\mathcal{P}\int_{-\infty}^{\infty}dz'\frac{{\rm Im}\left[f(z')\right]}{z'-z},
\end{alignat}
which gives, 
\begin{alignat}{1}
{\rm Re}\,\Sigma & _{\downarrow}(\mathbf{k},\omega)=\nonumber \\
 & \int\frac{d^{3}q}{(2\pi)^{3}}\frac{d\epsilon}{2\pi}\bigg[-{\rm Im}\Gamma(\mathbf{q},\epsilon){\rm Re}G_{\uparrow}^{(0)}(\mathbf{q}-\mathbf{k},\epsilon-\omega^{+})b(\epsilon)\nonumber \\
 & +{\rm Re}\Gamma(\mathbf{q},\epsilon){\rm Im}G_{\uparrow}^{(0)}(\mathbf{q}-\mathbf{k},\epsilon-\omega^{+})f(\epsilon-\omega)\bigg].
\end{alignat}

The above procedure for calculating the impurity Green's function
and vertex function breaks down for either a critical interaction
strength, temperature, or impurity concentration, when there exists
a tightly bound molecular state. For a balanced gas this is the condensation
of spontaneously created molecules and is the Thouless criterion for
superfluidity \cite{nozieres1985bose}, 
\begin{equation}
\Gamma^{-1}\left(\mathbf{q}=0,i\nu_{n}=0\right)=0,
\end{equation}
In this work we only consider the regimes away from the respective
molecular transitions.

\section{Virial expansion of an imbalanced Fermi gas}

\label{app:thermo}

Following the derivation of the imbalanced thermodynamic potential
by Refs.~\cite{liu2010_imb,Liu2013}, we write it in the following
third order form, 
\begin{alignat}{1}
\Omega=\Omega^{(1)}-k_{B}T\frac{2}{\lambda^3}\left[z_{\uparrow}z_{\downarrow}\Delta b_{2}+\frac{z_{\uparrow}^{2}z_{\downarrow}+z_{\uparrow}z_{\downarrow}^{2}}{2}\Delta b_{3}\right],
\end{alignat}
where $\Omega^{(1)}=\Omega^{(1)}(\mu_{\uparrow})+\Omega^{(1)}(\mu_{\downarrow})$
are the thermodynamic potential of a non-interacting Fermi gas for
each spin component and the thermal wavelength is $\lambda=\sqrt{2\pi/mk_{\rm B}T}$. The second order virial coefficient $\Delta b_{2}$
can be straight forwardly calculated \cite{beth}, and unitarity is given by
$\Delta b_2 = 1/\sqrt{2}$. The third order
virial coefficient $\Delta b_{3}$ can be found through a summation
over energies of the scattered states \cite{Liu2009,Rakshit2012}
or through field theoretical method \cite{Bedaque2003,Leyronas2011,Kaplan2011}, and at unitarity $\Delta b_3 = -0.35501\dots$.

With the virial expansion of the thermodynamic potential we can find
the density of each spin component, $n_{\sigma}=-\partial\Omega/\partial\mu_{\sigma}$.
We solve the majority chemical potential as in the finite temperature
$T$-matrix calculation, from the ideal gas at the same temperature,
and calculate the minority for a given density $x=n_{\downarrow}/n_{\uparrow}$.
In 3D $\tau=T/T_{{\rm F}}$ and $T_{{\rm F}}=\hbar^{2}(6\pi^{2}n)^{2/3}/2m/k_{B}$
is the Fermi temperature and we can find the dimensionless density
$\tilde{n}=n\lambda^{3}/2=4/(6\sqrt{\pi}\tau^{3/2})$ 
\begin{alignat}{1}
\tilde{n}_{\uparrow} & =\tilde{n}_{\uparrow}^{(1)}(z_{\uparrow})\\
\tilde{n}_{\downarrow} & =\tilde{n}_{\downarrow}^{(1)}(z_{\uparrow})+z_{\uparrow}z_{\downarrow}2\Delta b_{2}+\left(z_{\uparrow}^{2}z_{\downarrow}+2z_{\uparrow}z_{\downarrow}^{2}\right)\Delta b_{3},
\end{alignat}
where the ideal density is given by $\tilde{n}_{\sigma}^{(1)}(z)=(2/\sqrt{\pi})\int_{0}^{\infty}\sqrt{t}\left[ze^{-t}/(1+ze^{-t})\right]dt$.

\section{Virial expansion of Green's function}

\label{app:Green}

The virial expansion can be used to expand the self-energy in orders
of the non-interacting Green's function in powers of the fugacity
\cite{Leyronas2011,Hu2010prl,nishida2013,sun2015,Barth2014,Barth2015,vudtiwat2015,sun2017prl}.
For our highly imbalanced system we will utilize the fact that the
minority component chemical potential will always be large and negative
and the fugacity will be small. To begin then, we will only look at
diagrams with $O(z_{\downarrow}^{0})$ within the usual expansion
of the self-energy.

\begin{figure}
\centering{}\includegraphics{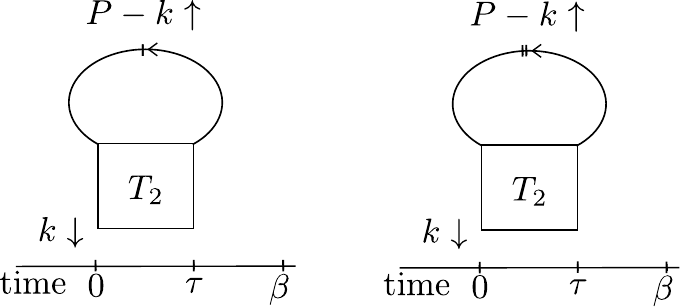} \caption{ (a) The lowest order contribution to the self-energy for the spin
$\downarrow$ minority Green's function, where $T_{2}$ is the two-body
T-matrix. (b) The third order, $O(z_{\downarrow}^{2})$, contribution
where the hole line is double slashed. \label{fig:sig_2}}
\end{figure}

\subsection{Second-order self energy}

The first term to contribute to the $\downarrow$ minority Green's
function is shown in Fig.~\ref{fig:sig_2}(a). The analytical expression
for the above diagram is given by: 
\begin{alignat}{1}
\Sigma_{\downarrow}^{(1)}(k,\tau)=z_{\uparrow}\int\frac{d\mathbf{P}}{(2\pi)^{3}}e^{\mu_{\downarrow}\tau}e^{-(\beta-\tau)\xi_{\mathbf{P}-\mathbf{k}}}T_{2}(P,\tau),
\end{alignat}
and the two-body T-matrix, $T_{2}(P,\tau)=e^{-\tau P^{2}/4m}T_{2}(0,\tau)$,
is the inverse Laplace transform of the two-body T-matrix, $t_{2}(s)=(4\pi/m)\left[a^{-1}-\sqrt{-ms}\right]^{-1}$,
\begin{alignat}{1}
T_{2}(0,\tau)=\int_{C_{\gamma}}\frac{ds}{2\pi i}e^{-\tau s}t_{2}(s).
\end{alignat}
The inverse Laplace transform is defined as a contour integral on
the Bromwich contour, which is a straight line in the complex plane
parallel to the imaginary axis and such that all of the function is
analytic to the left of the contour. It is clear from the definition
of $t_{2}(s)$ that there is a branch cut for all positive $s$ and
so we can take the Bromwich contour to be the positive real axis,
and if $a^{-1}<0$ include the additional contribution from the residue
due to the bound state contribution.

So we have, 
\begin{alignat}{1}
T_{2}(0,\tau)=-\Theta(a^{-1})\frac{8\pi e^{E_{b}\tau}}{m^{2}a}-\frac{4}{m^{3/2}}\int_{0}^{\infty}dxe^{-\tau x}\frac{\sqrt{x}}{x+E_{b}},
\end{alignat}
where $E_{b}=1/(ma^{2})$. Combining all of the above and taking the
Fourier transform for the imaginary time to Matsubara frequencies
we arrive at to $O(z_{\downarrow}^{0})$ 
\begin{alignat}{1}
\Sigma_{\downarrow}^{(1)} & (k,i\omega_{n})\nonumber \\
 & =z_{\uparrow}\int_{0}^{\infty}d\tau e^{i\omega_{n}\tau}\int\frac{d\mathbf{P}}{(2\pi)^{3}}e^{\mu_{\downarrow}\tau}e^{-(\beta-\tau)\xi_{\mathbf{P}-\mathbf{k}}}T_{2}(P,\tau),\nonumber \\
 & =z_{\uparrow}F(k,i\omega_{n}+\mu_{\downarrow}),
\end{alignat}
where, 
\begin{alignat}{1}
F(k,\omega^{+}) & =\int\frac{d\mathbf{P}}{(2\pi)^{3}}e^{-\beta\frac{(\mathbf{P}-\mathbf{k})^{2}}{2m}}\left[\int_{0}^{\infty}dx\frac{\rho_{2}(x)}{E^{+}-\frac{P^{2}}{4m}-x+\frac{(\mathbf{P}-\mathbf{k})^{2}}{2m}}+\frac{8\pi}{m^{2}a}\Theta(a^{-1})\frac{1}{\omega-\frac{P^{2}}{4m}-E_{b}+\frac{(\mathbf{P}-\mathbf{k})^{2}}{2m}}\right],\\
\end{alignat}

We have analytically continued the Matsubara frequencies to the real
axis as there are no more poles in the upper-complex plane. There
is an additional higher order, $O(z_{\downarrow})$, contribution
to $\Sigma_{\downarrow}^{(1)}(\mathbf{k},\omega)$ \cite{sun2015}
and we omit its contribution here.

\begin{figure}[t!]
\centering{}\includegraphics{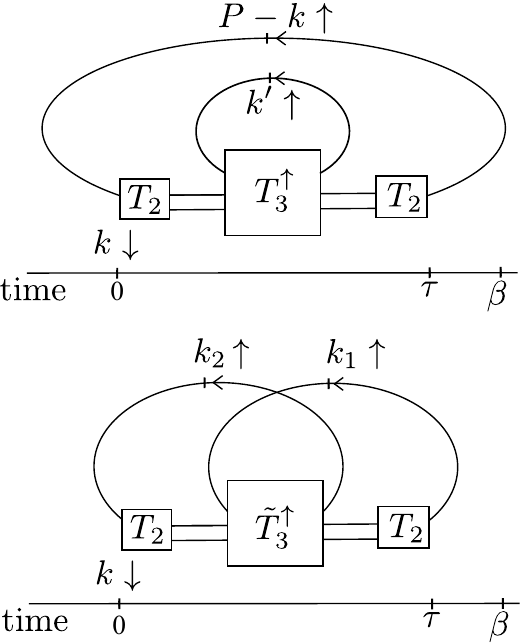} \caption{ The two diagrams for the spin $\downarrow$ self-energy, where $T_{3}$
is the three-body T-matrix and $T_{2}$ is the two-body T-matrix.
\label{fig:sig_3}}
\end{figure}

\subsection{Third-order self energy}

For the third order contributions we follow the calculation of the
diagrams from Refs.~\cite{sun2015,sun2017prl}, where for a two-component
Fermi gas there are six diagrams which contribute to the third order
self-energy. For every slashed line we have a power of fugacity (and
there will be an additional $e^{\mu_{\sigma}\tau}$ with the three
body STM-equations $T_{3}^{\sigma}$), we can see that to zeroth order
in the minority fugacity, only the diagrams in Fig.~\ref{fig:sig_3}(a)
and (b) will contribute. There is also an additional contribution
at the same order of the fugacity by slashing the second order diagram
twice that is found in Fig.~\ref{fig:sig_2}(b).

The first contribution to the third order self energy we calculate
is the double slashed diagram in Fig.~\ref{fig:sig_2}(b). The contribution
is obtained from $\Sigma^{(1)}$ by changing $G^{(0,1)}$ to $G^{(0,2)}$
and multiplying by a factor of $z_{\uparrow}$. The analytically continued
contribution is given by, 
\begin{alignat}{1}
\Sigma & _{\downarrow}^{(2,1)}(k,\omega^{+})\nonumber \\
 & =-z_{\uparrow}^{2}\int\frac{d\mathbf{P}}{(2\pi)^{3}}\int dx\frac{e^{-\beta\frac{(\mathbf{P}-\mathbf{k})^{2}}{m}}\rho_{2}(x)}{\omega^{+}+\mu_{\downarrow}-\left(\frac{P^{2}}{4m}+x+\frac{(\mathbf{P}-\mathbf{k})^{2}}{2m}\right)},
\end{alignat}
where again we have analytically continued to the real axis. The third
order diagrams in Fig.~\ref{fig:sig_3} give the following contributions
\cite{sun2015}, 
\begin{alignat}{1}
\Sigma_{\downarrow}^{(2,2)}(k,\omega^{+})= & z_{\uparrow}^{2}\int\frac{d\mathbf{P}d\mathbf{p}_{1}}{(2\pi)^{6}}\int_{0}^{\infty}dx\rho_{3}(\mathbf{p}_{1},\mathbf{p}_{1};x)\frac{e^{-\beta\left[\frac{(3\mathbf{p}_{1}+\mathbf{P}+\frac{(\mathbf{P}-\mathbf{k})^{2}}{2m})^{2}}{8m}\right]}}{\omega^{+}+\mu_{\downarrow}+\frac{P^{2}}{6m}+\frac{(\mathbf{P}-\mathbf{k})^{2}}{2m}-\frac{3(\mathbf{p}_{1}+\mathbf{P})^{2}}{8m}}\nonumber \\
+ & z_{\uparrow}^{2}\int\frac{d\mathbf{P}d\mathbf{p}_{1}}{(2\pi)^{6}}\int_{0}^{\infty}dx\tilde{\rho}_{3}(\mathbf{p}_{1},\mathbf{p}_{2};x)\frac{e^{-\beta\left[\frac{(3\mathbf{p}_{1}+\mathbf{P}+\frac{(\mathbf{P}-\mathbf{k})^{2}}{2m})^{2}}{8m}\right]}}{\omega^{+}+\mu_{\downarrow}+\frac{P^{2}}{6m}+\frac{(\mathbf{P}-\mathbf{k})^{2}}{2m}-\frac{3(\mathbf{p}_{1}+\mathbf{P})^{2}}{8m}}
\end{alignat}
where we define 
\begin{alignat}{1}
\rho_{3}(\mathbf{p}_{1},\mathbf{p}_{2};x)=\frac{1}{2\pi i}\biggl[ & t_{2}\left(x+i\delta-\frac{3p_{1}^{2}}{4m}\right)t_{3}(\mathbf{p}_{1},\mathbf{p}_{2};x+i\delta)t_{2}\left(x+i\delta-\frac{3p_{2}^{2}}{4m}\right)\nonumber \\
 & -t_{2}\left(x-i\delta-\frac{3p_{1}^{2}}{4m}\right)t_{3}(\mathbf{p}_{1},\mathbf{p}_{2};x-i\delta)t_{2}\left(x-i\delta-\frac{3p_{2}^{2}}{4m}\right)\biggl],
\end{alignat}
and $\tilde{\rho}_{3}$ is defined where the three-body T-matrix $t_{3}$
has removed a non one-particle irreducible contribution, 
\begin{alignat}{1}
\tilde{t}_{3}(\mathbf{p}_{1},\mathbf{p}_{2};s)=t_{3}(\mathbf{p}_{1},\mathbf{p}_{2};s)-\frac{1}{\frac{p_{1}^{2}+p_{2}^{2}}{m}+\frac{\mathbf{p}_{1}\cdot\mathbf{p}_{2}}{m}-s}.
\end{alignat}

The three-body integral equations $t_{3}(\mathbf{p}_{1},\mathbf{p}_{2};s)$
are defined in Appendix \ref{app:STM}. In the numerical calculation
we add a small imaginary part $\delta$ to deal with the poles and
branch cuts in the STM equations, we find this gives a small, but
negligible, shift to the final contribution to the self-energy. In
total to third order, for an imbalanced gas, we can see that the third
order contribution to the self energy is, 
\begin{alignat}{1}
\Sigma_{\downarrow}^{(3)}(k,\omega)=\Sigma_{\downarrow}^{(2,1)}(k,\omega)+\Sigma_{\downarrow}^{(2,2)}(k,\omega)
\end{alignat}

\section{Three body-integral equations}

\label{app:STM}

In the calculation of the self-energy to third order, i.e. $O(z_{\uparrow}^{2})$,
we need to calculate the vacuum three-body $T_{3}$ matrix, which
can be found from the Skornaikov-Ter Martirosian (STM) integral equation
\cite{STM}. Following the standard approach of the diagrammatic three-body
scattering $T_{3}$ matrix for an $\uparrow\uparrow\downarrow$ system~\cite{Bedaque1999,Brodsky2006,Leyronas2011,vudtiwat2015},
we have 
\begin{alignat}{1}
T_{3} & (p_{1},p_{2};P)=-G_{\downarrow}(P-p_{1}-p_{2})-\nonumber \\
 & \sum_{q}G_{\uparrow}(q)G_{\downarrow}(P-p_{1}-q)T_{2}(P-q)T_{3}(q,p_{2};P).
\end{alignat}
where we define the Green's function $G(q)=1/(q_{0}-\mathbf{q}^{2}+i0^{+})$
for four vector $q\equiv(\mathbf{q},q_{0})$ and $\sum_{q}\equiv\int_{\mathbf{q},q_{0}}$.
Performing the $q_{0}$ integral, changing the coordinates, going
to the center-of-mass frame, and using the on-shell energies we can
simplify the three-body $T$ matrix to $T_{3}(\{\mathbf{p}_{1},\varepsilon_{\mathbf{p}_{1}}\},\{\mathbf{p}_{2},\varepsilon_{\mathbf{p}_{2}}\};\{\mathbf{P}=0,s\})\equiv t_{3}(\mathbf{p}_{1},\mathbf{p}_{2};s)$.
This gives in total the integral equation, 
\begin{alignat}{1}
t_{3}(\mathbf{p},\mathbf{k};s)=\frac{1}{s-\frac{k^{2}+p^{2}}{m}-\frac{\mathbf{k}\cdot\mathbf{p}}{m}}+\int\frac{d\mathbf{q}}{(2\pi)^{3}}\frac{t_{2}\left(s-\frac{3q^{2}}{4m}\right)}{s-\frac{k^{2}+p^{2}}{m}-\frac{\mathbf{k}\cdot\mathbf{p}}{m}}t_{3}(\mathbf{q},\mathbf{k};s),
\end{alignat}
where $s$ is the total center-of-mass energy 
We can decompose the STM equations into angular momentum channels,
where 
\begin{alignat}{1}
t_{3}(\mathbf{p},\mathbf{k};s) & =\sum_{l}(2l+1)P_{l}(x)t_{3}^{(l)}(p,k,s),\\
t_{3}^{(l)}(p,k,s)\  & =\frac{1}{2}\int_{-1}^{1}dxP_{l}(x)t_{3}(\mathbf{p},\mathbf{k},s),
\end{alignat}
where $x=\cos\left(\hat{\mathbf{p}}\cdot\hat{\mathbf{k}}\right)$
and $P_{l}(x)$ is the Legendre polynomials. The decoupled STM equations
for the angular momentum channels is, 
\begin{alignat}{1}
t_{3}^{(l)}(p,k,s)=\frac{m}{pk}Q_{l}\left[\frac{m}{pk}\left(s-\frac{p^{2}}{m}-\frac{k^{2}}{m}\right)\right]+\int\frac{dq}{2\pi^{2}}q^{2}t_{2}\left(s-\frac{3q^{2}}{4m}\right)\frac{m}{pq}Q_{l}\left[\frac{m}{pq}\left(s-\frac{p^{2}}{m}-\frac{q^{2}}{m}\right)\right]t_{3}^{(l)}(p,q,s),
\end{alignat}
and $Q_{l}(z)=\frac{1}{2}\int_{-1}^{1}dx\frac{1}{z-x}P_{l}(x)$ is
the Legendre function of the second kind. In the numerical calculations
we take 10 angular momentum channels and find that the final results
are independent on the number of channels.

\bibliography{polaron_paper}
 

\end{document}